\documentclass[12pt]{iopart}
\usepackage{latexsym}
\usepackage{fullpage}
\usepackage{graphics}
\usepackage{graphicx}

\begin{document}
\title{Swift Pointing and Gravitational-Wave Bursts from 
Gamma-Ray Burst Events}

\author{Patrick J. Sutton,\footnote{Also Center for Gravitational Physics and
Geometry and Department of Physics; \texttt{e-mail:psutton@gravity.psu.edu}}~ 
Lee Samuel Finn\footnote{Also Center for
Gravitational Physics and Geometry, Department
of Physics, and Department of Astronomy and Astrophysics;
\texttt{e-mail:lsf@gravity.psu.edu}}} 
\address{Center for Gravitational Wave Physics, Department of
Physics,\\The Pennsylvania State University, University Park, PA
16802, USA }
\author{Badri Krishnan\footnote{Also Center for Gravitational Wave
Physics and Center for Gravitational Physics and Geometry,
Pennsylvania State University, University Park, PA-16801, USA; 
\texttt{e-mail:badkri@aei.mpg.de}}}
\address{Max Planck Institut f\"ur Gravitationsphysik,\\ Am M\"uhlenberg
1, 14476 Golm, Germany}

\begin{abstract}
The currently accepted model for gamma-ray burst phenomena involves
the violent formation of a rapidly rotating solar-mass
black hole. Gravitational waves should be associated with the black-hole 
formation, and their detection would permit this model to be
tested.
Even upper limits on the gravitational-wave strength associated 
with gamma-ray bursts could constrain the gamma-ray burst model.
This requires joint observations of gamma-ray
burst events with gravitational and gamma-ray detectors. Here we
examine how the quality of an upper limit on the gravitational-wave
strength associated with gamma-ray bursts depends on the
relative orientation of the gamma-ray-burst and gravitational-wave 
detectors, and apply our results to the particular case of the
Swift Burst-Alert Telescope (BAT) and the LIGO gravitational-wave
detectors. A result of this investigation is a science-based
``figure of merit'' that can be used, together with other
mission constraints, to optimize the pointing of the Swift telescope
for the detection of gravitational waves associated with gamma-ray
bursts. 
\end{abstract}

\section{Introduction}
\label{sec:Introduction}

The currently accepted model for gamma-ray burst phenomena
involves the violent formation of a rapidly rotating 
approximately solar-mass black hole surrounded by a similarly 
massive debris torus \cite{MeRe:93, ReMe:94}. 
A gravitational-wave burst is likely to be associated with the
formation of this ``central engine,'' and 
the observation of such a gravitational-wave burst may  
reveal details of the central engine that cannot be
revealed through observations of the gamma rays alone. 
In this paper we examine how 
joint observations from the 
LIGO gravitational-wave detectors \cite{Si:01} 
and the Swift gamma-ray burst satellite \cite{Swift} 
can be used to detect or place upper limits on 
gravitational-wave emission by gamma-ray burst events.

Finn, Mohanty, and Romano \cite{FiMoRo:99} 
have described how the cross-correlated output of
two gravitational-wave detectors, taken in coincidence with
gamma-ray burst (GRB) events, can be used to detect or place upper
limits on the emission of gravitational-wave bursts (GWBs) by GRBs. 
Finn et al.~estimate that 1000 GRB observations
combined with observations from the initial LIGO detectors could
produce an upper limit on the gravitational-wave strain associated
with GRBs of approximately $h_{\mbox{\tiny{}RMS}} \le
1.7\times10^{-22}$ at 95\% confidence.

In their original work Finn et al.~
assumed that GRBs would be detected isotropically; i.e., 
that the GRB detector had an isotropic antenna pattern.
They did note, however,
that the Swift satellite \cite{Swift}, 
a next-generation multi-wavelength satellite
dedicated to the study of GRBs, does not have an isotropic antenna
pattern and that this has potentially important consequences for the
ability of the combined GRB/GWB detector
array to detect or limit the gravitational-wave flux on Earth owing to
GRBs.  Here we study this question specifically in the context of the
Swift satellite and the LIGO gravitational-wave detectors; i.e.,~we
determine, as a function of Swift's pointing, the sensitivity
of the Swift/LIGO detector array to gravitational waves from GRBs, and
propose a figure-of-merit that can be used in Swift mission scheduling
to optimize the sensitivity of the Swift/LIGO array to the
gravitational-wave flux from GRBs.  We find that the upper limit that
can be placed on $h_{\mbox{\tiny{}RMS}}$ differs by a factor of 2
between best and worst orientations of the satellite.

We begin in Section~2 with a review of how one can 
detect place 
upper limits on the gravitational-wave strength associated with GRBs.
In Section~3 we extract the direction dependence of this upper limit 
and apply to the case of Swift.  We conclude with some brief remarks 
in Section~4.

\section{Observing a GRB--GWB Association}

In this section we briefly review the analysis presented in \cite{FiMoRo:99}.

Consider a set of $N$ GRB detections,  each characterized by the 
direction to the source $\widehat{\Omega}_k$ and the arrival time 
${\tau}_k$ of the burst at Earth's barycenter.  A plane gravitational wave
incident on the detector pair from the direction $\widehat{\Omega}_k$
will lead to correlated detector responses with a time lag equal to 
\begin{equation}
\Delta t_k = t^{(2)}_k - t^{(1)}_k \, , 
\end{equation}
where $t^{(i)}_k$ is the arrival time of the burst at detector $i$,
which depends only on $\tau_k$, $\widehat{\Omega}_k$, 
and the detector location.

Let $s_i(t)$ be the output of gravitational wave detector
$\mathcal{D}_i$, which we assume to consist of detector noise $n_i(t)$
and a possible gravitational wave signal $h_i(t)$ produced by the
GRB source:
\begin{equation}\label{eq:s} 
s_i(t) = n_i(t) + h_i(t) \, .
\end{equation}
Finn et al.~\cite{FiMoRo:99} define 
\begin{equation}
\label{eq:cross} 
S(\widehat{\Omega}_k,\tau_k) 
  =  \left<s_1, s_2\right> 
 :=  \int_0^T \!\!dt \int_0^T \!\!dt' \, s_{1}(t_k^{(1)}-t)
         \,Q(t-t')\, s_{2}(t_k^{(2)}-t') \, ,
\end{equation}
as the energy in the cross-correlation of the
two detectors corresponding to the GRB characterized by
$(\widehat{\Omega}_k,\tau_k)$. 
Here $Q$ is a freely specifiable symmetric filter function, 
and $T$ is chosen large enough to encompass the
range of possible times by which the gravitational waves from a GRB
event may precede the gamma rays, which is typically thought to be of
order 1 s for GRBs produced by internal shocks and 100 s for GRBs
produced by external shocks \cite{SaPi:97, KoPiSa:97}.

The presence of a GWB in the data will increase the mean 
value of the cross-correlation (\ref{eq:cross}) over its 
mean value when no GWBs are present. 
Finn et al.~\cite{FiMoRo:99} showed that this effect can be used to 
detect 
or place upper limits on the gravitational-wave strain associated 
with GRBs by 
comparing the mean value $S_{\mbox{\tiny{}on}}$ 
of the cross-correlation statistic 
for a set of GRB observations to the mean cross-correlation 
$S_{\mbox{\tiny{}off}}$ 
computed using random $(\widehat{\Omega},\tau)$ not associated
with any GRB.   
The expected difference between  $S_{\mbox{\tiny{}on}}$ and 
$S_{\mbox{\tiny{}off}}$ is simply the contribution 
due to GWBs from the GRB events; using (\ref{eq:s}), 
we have 
\begin{equation}
\label{eq:hcross}
S_{\mbox{\tiny{}on}} - S_{\mbox{\tiny{}off}}  
~=~ \overline{\left<h_1, h_2\right>}
~=~ \int_0^T \!\!dt \int_0^T \!\!dt' \, Q(t-t')  \, 
         \overline{h_{1}(t_k^{(1)}-t) \, h_{2}(t_k^{(2)}-t')} 
         \, , 
\end{equation}
where the overline denotes an average over the population of GRBs.
Since the right-hand side of (\ref{eq:hcross}) is quadratic in the 
gravitational-wave strain, a measurement of or limit on 
the difference in cross-correlations leads directly to an 
estimate of or upper limit on the gravitational-wave strain 
associated with GRBs.

Here we are not interested in the absolute value of the upper 
limit that can be achieved, but rather in how that upper limit 
varies according to the relative orientation of the GWB/GRB detector array. 
We shall extract this dependence in the next section.

\section{Direction-dependence of the upper limit}

The analysis described in \cite{FiMoRo:99} assumed that the arms of
the two LIGO gravitational-wave detectors all reside  in the same
plane, that pairs of arms are parallel to each other, and that the
antenna pattern of the detectors is isotropic on the sky. In this
section we relax all of these approximations; i.e., we properly
account for the position and orientation of the two LIGO detectors on
the Earth and the dependence of their sensitivity to the direction to
the GRB source. Our result is an expression for the dependence of the
upper limit on the population-averaged gravitational-wave strength
$\overline{\left<h_1,h_2\right>}$ as a function of the distribution 
of \emph{detected} GRBs on the sky. 
We also combine this result with
the directional sensitivity of the Swift detector to determine the
dependence on Swift pointing of the upper limit on
$\overline{\left<h_1,h_2\right>}$ that can be set by joint LIGO/Swift
observations.

The gravitational-wave component $h_i$ of the LIGO detector output 
is a linear function of the physical gravitational-wave strain
$h_{ab}(t,\vec{x})$,
\begin{equation}\label{eq:h}
h_i(t) = h_{ab}(t,\vec{x}_i)\, d^{ab}_i \, ,
\end{equation}
where $\vec{x}_i$ is the gravitational-wave detector's location 
and $d^{ab}_i$ is the detector response function.  For
interferometer $i$ with arms pointing in the
directions $\widehat{X}_i$, $\widehat{Y}_i$, the 
latter is 
\begin{equation}\label{eq:dIFO}
d^{ab}_i = \frac{1}{2} ( \widehat{X}^a_i \widehat{X}^b_i
  - \widehat{Y}^a_i \widehat{Y}^b_i )\, .
\end{equation}

It is convenient to decompose the gravitational wave 
into its two polarization states,
\begin{equation}\label{hdecomp}
h_{ab}(t)  =  h_+(t) \epsilon^+_{ab}(\widehat{\Omega})
              + h_+(t) \epsilon^\times_{ab}(\widehat{\Omega}) \, .
\end{equation}
See \cite{AlRo:99} for one choice of the polarization 
tensors $\epsilon^A_{ab}(\widehat{\Omega})$.
Lacking any detailed model for the gravitational waves that may
be produced in a GRB event, we assume that 
(i) the waves have equal power in the two polarizations 
($\overline{h_+(t) h_+(t')} = \overline{h_\times(t) h_\times(t')}$)
and (ii) the two polarizations are uncorrelated
($\overline{h_+(t) h_\times(t')} = 0$).

Focus attention now on the mean gravitational-wave contribution
$\overline{\left<h_1,h_2\right>}$ to the cross-correlation 
statistic (\ref{eq:cross}). For GWBs
arriving from the direction $\widehat{\Omega}$, it can be shown that
\begin{equation}
\overline{\left<h_1,h_2\right>} 
  =  \rho_{\mbox{\tiny{}GWB}}(\widehat{\Omega}|d_1,d_2)
     \int_0^T\!\!dt\! \int_0^T\!\!dt'\,Q(t-t')
     \,\overline{\left<h_{+}(t)\,h_{+}(t')\right>}\,,
\end{equation}
where 
\begin{equation}\label{eq:rhoLIGO} 
\rho_{\mbox{\tiny{}GWB}}(\widehat{\Omega}|d_1,d_2)
  \equiv  \sum_{A=+,\times}  
          d_1^{ab} \epsilon^A_{ab}(\widehat{\Omega}) \, 
          d_2^{cd} \epsilon^A_{cd}(\widehat{\Omega}) 
\end{equation}
describes the direction-dependence of the sensitivity of the 
gravitational-wave detector pair to the GWB.

To complete the evaluation of $\overline{\left<h_1,h_2\right>}$ consider 
to the fraction of GRB detections that arise from different
patches on the sky. Since the intrinsic GRB population is isotropic,
the distribution of detection on the sky depends entirely on the
directional sensitivity of the GRB detector. Let the fraction of 
GRB detections in a sky patch of area $d^2\Omega$ centered at
$\widehat{\Omega}$ be given by 
\begin{equation}
\rho_{\mbox{\tiny{}GRB}}(\widehat{\Omega}|\widehat{\Omega}',\widehat{n})
d^2\widehat{\Omega}\,,
\end{equation}
where $\widehat{\Omega}'$ is 
the direction in which the GRB detector is pointed, and $\widehat{n}$  
describes the rotation of the satellite about its pointing direction.
It can then be shown that the upper limit on the squared 
gravitational-wave strain averaged over the observed GRB population 
when the orientation of GWB and the GRB detectors are
given by $(d_1, d_2, \widehat{\Omega}, \widehat{n})$ is
inversely proportional to 
\begin{equation}\label{eq:FOM}
  \zeta(\widehat{\Omega}, \widehat{n},d_1, d_2)
  =  
  \int \!d^2\widehat{\Omega}'\,
  \rho_{\mbox{\tiny{}GRB}}(\widehat{\Omega}'|\widehat{\Omega},\widehat{n})\,
  \rho_{\mbox{\tiny{}GWB}}(\widehat{\Omega}'|d_1, d_2) 
  \, .
\end{equation}


Clearly $\zeta$ can be regarded as a figure of merit that describes
how capable the gravitational-wave/gamma-ray burst detector
combination is at identifying GWBs associated with GRBs as a function
of the detector orientations. This figure of merit may be normalized
to have a maximum of unity; however, regardless of the normalization
\begin{equation}
  \zeta(\widehat{\Omega}',\widehat{n}', d_1, d_2) /
  \zeta(\widehat{\Omega},\widehat{n}, d_1, d_2) 
\end{equation}
is the ratio of the upper limits on the squared gravitational-wave
amplitude that can be attained by orienting the GRB satellite as
$(\widehat{\Omega},\widehat{n})$ versus
$(\widehat{\Omega}',\widehat{n}')$. 
To the extent that, e.g., the GRB detector orientation 
$(\widehat{\Omega},\widehat{n})$ can be manipulated on orbit, 
choosing orientations that maximize $\zeta$ will lead to larger 
signal contributions (\ref{eq:hcross}) to the cross-correlation 
and thus more sensitive measurements of the
gravitational-wave strength associated with GRBs.

Let us now consider the special case of the Burst-Alert Telescope
(BAT) on the Swift satellite \cite{Swift} and the LIGO gravitational
wave detectors \cite{Si:01}.  The BAT is a wide field-of-view
coded-aperture gamma-ray imager that will detect and locate GRBs with
arc-minute positional accuracy. Its sensitivity to GRBs depends on the
the angle $\lambda$ between the line of sight to the GRB and the BAT
axis, as well as the rotational orientation of the satellite about 
the BAT axis.  
Averaged over this azimuthal angle, the BAT sensitivity as a function
of $\lambda$ is approximately\footnote{
See the BAT section of the Swift homepage,  
http://swift.gsfc.nasa.gov/science/instruments/bat.html  
} 
\begin{equation}\label{eq:rhoSwift}
\rho_{\mbox{\tiny{}Swift}}
  =  \left\{
       \begin{array}{ll}
         2\cos\lambda -1 + 0.077\sin{[13(1-\cos\lambda)]} & 
         \lambda \in [0,\pi/3] \, , \\
         0 & \mbox{otherwise} \, . 
       \end{array}
     \right.
\end{equation}
For the purposes of illustration we will
use this azimuthal-angle averaged expression for the BAT sensitivity. 

\begin{figure} 
  \begin{center}
  \includegraphics[height=8cm]{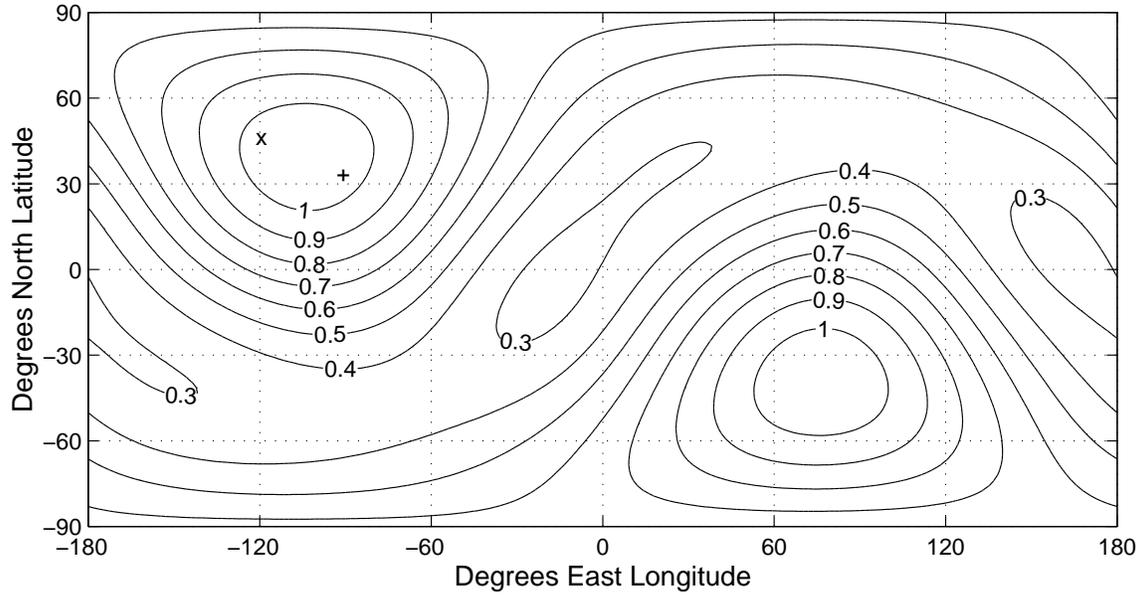}
  \caption{
\label{fig:FOM} 
Figure of merit $\zeta$ (\ref{eq:FOM}) for Swift pointing in Earth-based
coordinates, produced by convolving the LIGO sensitivity pattern
$\rho_{\mbox{\tiny{}GWB}}$ (\ref{eq:rhoLIGO}) with the Swift sensitivity
function $\rho_{\mbox{\tiny{}Swift}}$ (\ref{eq:rhoSwift}).  The 
figure of merit is nowhere zero, having a range of $[0.25,1.00]$ and an
all-sky average of 0.56.
The $+$ and $\times$ mark the locations of the LIGO Livingston 
and LIGO Hanford detectors.
}
  \end{center}
\end{figure}

Convolving $\rho_{\mbox{\tiny{}GWB}}$ (\ref{eq:rhoLIGO}) for the LIGO 
detector array (see \cite{AlHaJoLaWe:00}) with the Swift sensitivity
function $\rho_{\mbox{\tiny{}Swift}}$ as in equation (\ref{eq:FOM}) gives the
figure of merit $\zeta$ for the Swift pointing, which is shown in
Figure~\ref{fig:FOM}. The figure of merit is nowhere zero, varying
by a factor of approximately 4 between best (near zenith of detectors)
and worst (near planes of detectors at 45 degrees from arms)
orientations of Swift; this translates into a factor 2 difference 
in amplitude sensitivity.  The all-sky average of the figure of merit is
0.56 times the maximum value.

\section{Conclusion}

In this article, we have evaluated how the quality of an upper limit on the
gravitational-wave strength associated with gamma-ray burst
observations depends on the relative orientation of the gamma-ray
burst and gravitational-wave detectors, with particular application to the
Swift Burst-Alert Telescope (BAT) and the LIGO gravitational wave
detectors. Setting aside other physical and science constraints on the
Swift mission, careful choice of BAT pointing leads to an upper limit
on the observed GRB population-averaged mean-square gravitational-wave
strength a factor of two lower than the upper limit resulting from
pointing that does not take this science into account.

\section{Acknowledgments}

We are grateful to Margaret Chester, Shiho Kobayashi, 
and John Nousek for useful discussions.
LSF and PJS acknowledge NSF grant PHY~00-99559, LSF, PJS and BK
acknowledge the Center for Gravitational Wave Physics, which is
supported by the NSF under cooperative agreement PHY~01-14375, and BK
acknowledges the support of the Albert Einstein Institut.




\end{document}